\documentclass[conference,10pt]{IEEEtran}
\IEEEoverridecommandlockouts

\usepackage{cite}
\usepackage[T1]{fontenc}
\usepackage{graphicx}
\usepackage{amssymb}
\usepackage{amsmath}
\usepackage{amsthm}
\usepackage{microtype}
\usepackage{balance}
\usepackage{subfigure}
\usepackage{xcolor} 
\usepackage{booktabs}

\usepackage{amsfonts}
\usepackage{algorithm}
\usepackage{algpseudocode}
 
\usepackage{graphicx}
\usepackage{textcomp}
\usepackage{xcolor}
\usepackage{microtype}
\usepackage[caption=false,font=footnotesize]{subfig}
\usepackage{adjustbox}
\usepackage{balance}

\usepackage{tablefootnote}
\usepackage[caption=false,font=footnotesize]{subfig}

\usepackage{amssymb}
\usepackage{amsfonts}
\usepackage{mathrsfs}
\usepackage{xspace}
\usepackage{bm}
\usepackage{upgreek}

\newcommand{\safemath}[2]{\newcommand{#1}{\ensuremath{#2}\xspace}}

\safemath{\bma}{\mathbf{a}}
\safemath{\bmb}{\mathbf{b}}
\safemath{\bmc}{\mathbf{c}}
\safemath{\bmd}{\mathbf{d}}
\safemath{\bme}{\mathbf{e}}
\safemath{\bmf}{\mathbf{f}}
\safemath{\bmg}{\mathbf{g}}
\safemath{\bmh}{\mathbf{h}}
\safemath{\bmi}{\mathbf{i}}
\safemath{\bmj}{\mathbf{j}}
\safemath{\bmk}{\mathbf{k}}
\safemath{\bml}{\mathbf{l}}
\safemath{\bmm}{\mathbf{m}}
\safemath{\bmn}{\mathbf{n}}
\safemath{\bmo}{\mathbf{o}}
\safemath{\bmp}{\mathbf{p}}
\safemath{\bmq}{\mathbf{q}}
\safemath{\bmr}{\mathbf{r}}
\safemath{\bms}{\mathbf{s}}
\safemath{\bmt}{\mathbf{t}}
\safemath{\bmu}{\mathbf{u}}
\safemath{\bmv}{\mathbf{v}}
\safemath{\bmw}{\mathbf{w}}
\safemath{\bmx}{\mathbf{x}}
\safemath{\bmy}{\mathbf{y}}
\safemath{\bmz}{\mathbf{z}}
\safemath{\bmzero}{\mathbf{0}}
\safemath{\bmone}{\mathbf{1}}

\bmdefine{\biad}{a}
\bmdefine{\bibd}{b}
\bmdefine{\bicd}{c}
\bmdefine{\bidd}{d}
\bmdefine{\bied}{e}
\bmdefine{\bifd}{f}
\bmdefine{\bigd}{g}
\bmdefine{\bihd}{h}
\bmdefine{\biid}{i}
\bmdefine{\bijd}{j}
\bmdefine{\bikd}{k}
\bmdefine{\bild}{l}
\bmdefine{\bimd}{m}
\bmdefine{\bind}{n}
\bmdefine{\biod}{o}
\bmdefine{\bipd}{p}
\bmdefine{\biqd}{q}
\bmdefine{\bird}{r}
\bmdefine{\bisd}{s}
\bmdefine{\bitd}{t}
\bmdefine{\biud}{u}
\bmdefine{\bivd}{v}
\bmdefine{\biwd}{w}
\bmdefine{\bixd}{x}
\bmdefine{\biyd}{y}
\bmdefine{\bizd}{z}

\bmdefine{\bixid}{\xi}
\bmdefine{\bilambdad}{\lambda}
\bmdefine{\bimud}{\mu}
\bmdefine{\bithetad}{\theta}
\bmdefine{\biphid}{\phi}
\bmdefine{\bideltad}{\delta}

\safemath{\bmia}{\biad}
\safemath{\bmib}{\bibd}
\safemath{\bmic}{\bicd}
\safemath{\bmid}{\bidd}
\safemath{\bmie}{\bied}
\safemath{\bmif}{\bifd}
\safemath{\bmig}{\bigd}
\safemath{\bmih}{\bihd}
\safemath{\bmii}{\biid}
\safemath{\bmij}{\bijd}
\safemath{\bmik}{\bikd}
\safemath{\bmil}{\bild}
\safemath{\bmim}{\bimd}
\safemath{\bmin}{\bind}
\safemath{\bmio}{\biod}
\safemath{\bmip}{\bipd}
\safemath{\bmiq}{\biqd}
\safemath{\bmir}{\bird}
\safemath{\bmis}{\bisd}
\safemath{\bmit}{\bitd}
\safemath{\bmiu}{\biud}
\safemath{\bmiv}{\bivd}
\safemath{\bmiw}{\biwd}
\safemath{\bmix}{\bixd}
\safemath{\bmiy}{\biyd}
\safemath{\bmiz}{\bizd}

\safemath{\bmxi}{\bixid}
\safemath{\bmlambda}{\bilambdad}
\safemath{\bmmu}{\bimud}
\safemath{\bmtheta}{\bithetad}
\safemath{\bmphi}{\biphid}
\safemath{\bmdelta}{\bideltad}

\safemath{\bA}{\mathbf{A}}
\safemath{\bB}{\mathbf{B}}
\safemath{\bC}{\mathbf{C}}
\safemath{\bD}{\mathbf{D}}
\safemath{\bE}{\mathbf{E}}
\safemath{\bF}{\mathbf{F}}
\safemath{\bG}{\mathbf{G}}
\safemath{\bH}{\mathbf{H}}
\safemath{\bI}{\mathbf{I}}
\safemath{\bJ}{\mathbf{J}}
\safemath{\bK}{\mathbf{K}}
\safemath{\bL}{\mathbf{L}}
\safemath{\bM}{\mathbf{M}}
\safemath{\bN}{\mathbf{N}}
\safemath{\bO}{\mathbf{O}}
\safemath{\bP}{\mathbf{P}}
\safemath{\bQ}{\mathbf{Q}}
\safemath{\bR}{\mathbf{R}}
\safemath{\bS}{\mathbf{S}}
\safemath{\bT}{\mathbf{T}}
\safemath{\bU}{\mathbf{U}}
\safemath{\bV}{\mathbf{V}}
\safemath{\bW}{\mathbf{W}}
\safemath{\bX}{\mathbf{X}}
\safemath{\bY}{\mathbf{Y}}
\safemath{\bZ}{\mathbf{Z}}

\safemath{\bZero}{\mathbf{0}}
\safemath{\bOne}{\mathbf{1}}
\safemath{\bDelta}{\mathbf{\Delta}}
\safemath{\bLambda}{\mathbf{\UpLambda}}
\safemath{\bPhi}{\mathbf{\Upphi}}
\safemath{\bSigma}{\mathbf{\Upsigma}}
\safemath{\bOmega}{\mathbf{\Upomega}}
\safemath{\bTheta}{\mathbf{\Uptheta}}

\bmdefine{\biAd}{A}
\bmdefine{\biBd}{B}
\bmdefine{\biCd}{C}
\bmdefine{\biDd}{D}
\bmdefine{\biEd}{E}
\bmdefine{\biFd}{F}
\bmdefine{\biGd}{G}
\bmdefine{\biHd}{H}
\bmdefine{\biId}{I}
\bmdefine{\biJd}{J}
\bmdefine{\biKd}{K}
\bmdefine{\biLd}{L}
\bmdefine{\biMd}{M}
\bmdefine{\biOd}{N}
\bmdefine{\biPd}{O}
\bmdefine{\biQd}{P}
\bmdefine{\biRd}{R}
\bmdefine{\biSd}{S}
\bmdefine{\biTd}{T}
\bmdefine{\biUd}{U}
\bmdefine{\biVd}{V}
\bmdefine{\biWd}{W}
\bmdefine{\biXd}{X}
\bmdefine{\biYd}{Y}
\bmdefine{\biZd}{Z}

\bmdefine{\biDelta}{\Delta}
\bmdefine{\biLambda}{\Lambda}
\bmdefine{\biPhi}{\Phi}
\bmdefine{\biSigma}{\Sigma}
\bmdefine{\biOmega}{\Omega}
\bmdefine{\biTheta}{\Theta}

\safemath{\bimA}{\biAd}
\safemath{\bimB}{\biBd}
\safemath{\bimC}{\biCd}
\safemath{\bimD}{\biDd}
\safemath{\bimE}{\biEd}
\safemath{\bimF}{\biFd}
\safemath{\bimG}{\biGd}
\safemath{\bimH}{\biHd}
\safemath{\bimI}{\biId}
\safemath{\bimJ}{\biJd}
\safemath{\bimK}{\biKd}
\safemath{\bimL}{\biLd}
\safemath{\bimM}{\biMd}
\safemath{\bimN}{\biNd}
\safemath{\bimO}{\biOd}
\safemath{\bimP}{\biPd}
\safemath{\bimQ}{\biQd}
\safemath{\bimR}{\biRd}
\safemath{\bimS}{\biSd}
\safemath{\bimT}{\biTd}
\safemath{\bimU}{\biUd}
\safemath{\bimV}{\biVd}
\safemath{\bimW}{\biWd}
\safemath{\bimX}{\biXd}
\safemath{\bimY}{\biYd}
\safemath{\bimZ}{\biZd}

\safemath{\bimDelta}{\biDelta}
\safemath{\bimLambda}{\biLambda}
\safemath{\bimPhi}{\biPhi}
\safemath{\bimSigma}{\biSigma}
\safemath{\bimOmega}{\biOmega}
\safemath{\bimTheta}{\biTheta}

\safemath{\setA}{\mathcal{A}}
\safemath{\setB}{\mathcal{B}}
\safemath{\setC}{\mathcal{C}}
\safemath{\setD}{\mathcal{D}}
\safemath{\setE}{\mathcal{E}}
\safemath{\setF}{\mathcal{F}}
\safemath{\setG}{\mathcal{G}}
\safemath{\setH}{\mathcal{H}}
\safemath{\setI}{\mathcal{I}}
\safemath{\setJ}{\mathcal{J}}
\safemath{\setK}{\mathcal{K}}
\safemath{\setL}{\mathcal{L}}
\safemath{\setM}{\mathcal{M}}
\safemath{\setN}{\mathcal{N}}
\safemath{\setO}{\mathcal{O}}
\safemath{\setP}{\mathcal{P}}
\safemath{\setQ}{\mathcal{Q}}
\safemath{\setR}{\mathcal{R}}
\safemath{\setS}{\mathcal{S}}
\safemath{\setT}{\mathcal{T}}
\safemath{\setU}{\mathcal{U}}
\safemath{\setV}{\mathcal{V}}
\safemath{\setW}{\mathcal{W}}
\safemath{\setX}{\mathcal{X}}
\safemath{\setY}{\mathcal{Y}}
\safemath{\setZ}{\mathcal{Z}}
\safemath{\emptySet}{\varnothing}

\safemath{\colA}{\mathscr{A}}
\safemath{\colB}{\mathscr{B}}
\safemath{\colC}{\mathscr{C}}
\safemath{\colD}{\mathscr{D}}
\safemath{\colE}{\mathscr{E}}
\safemath{\colF}{\mathscr{F}}
\safemath{\colG}{\mathscr{G}}
\safemath{\colH}{\mathscr{H}}
\safemath{\colI}{\mathscr{I}}
\safemath{\colJ}{\mathscr{J}}
\safemath{\colK}{\mathscr{K}}
\safemath{\colL}{\mathscr{L}}
\safemath{\colM}{\mathscr{M}}
\safemath{\colN}{\mathscr{N}}
\safemath{\colO}{\mathscr{O}}
\safemath{\colP}{\mathscr{P}}
\safemath{\colQ}{\mathscr{Q}}
\safemath{\colR}{\mathscr{R}}
\safemath{\colS}{\mathscr{S}}
\safemath{\colT}{\mathscr{T}}
\safemath{\colU}{\mathscr{U}}
\safemath{\colV}{\mathscr{V}}
\safemath{\colW}{\mathscr{W}}
\safemath{\colX}{\mathscr{X}}
\safemath{\colY}{\mathscr{Y}}
\safemath{\colZ}{\mathscr{Z}}

\safemath{\opA}{\mathbb{A}}
\safemath{\opB}{\mathbb{B}}
\safemath{\opC}{\mathbb{C}}
\safemath{\opD}{\mathbb{D}}
\safemath{\opE}{\mathbb{E}}
\safemath{\opF}{\mathbb{F}}
\safemath{\opG}{\mathbb{G}}
\safemath{\opH}{\mathbb{H}}
\safemath{\opI}{\mathbb{I}}
\safemath{\opJ}{\mathbb{J}}
\safemath{\opK}{\mathbb{K}}
\safemath{\opL}{\mathbb{L}}
\safemath{\opM}{\mathbb{M}}
\safemath{\opN}{\mathbb{N}}
\safemath{\opO}{\mathbb{O}}
\safemath{\opP}{\mathbb{P}}
\safemath{\opQ}{\mathbb{Q}}
\safemath{\opR}{\mathbb{R}}
\safemath{\opS}{\mathbb{S}}
\safemath{\opT}{\mathbb{T}}
\safemath{\opU}{\mathbb{U}}
\safemath{\opV}{\mathbb{V}}
\safemath{\opW}{\mathbb{W}}
\safemath{\opX}{\mathbb{X}}
\safemath{\opY}{\mathbb{Y}}
\safemath{\opZ}{\mathbb{Z}}
\safemath{\opZero}{\mathbb{O}}
\safemath{\identityop}{\opI}

\safemath{\veca}{\bma}
\safemath{\vecb}{\bmb}
\safemath{\vecc}{\bmc}
\safemath{\vecd}{\bmd}
\safemath{\vece}{\bme}
\safemath{\vecf}{\bmf}
\safemath{\vecg}{\bmg}
\safemath{\vech}{\bmh}
\safemath{\veci}{\bmi}
\safemath{\vecj}{\bmj}
\safemath{\veck}{\bmk}
\safemath{\vecl}{\bml}
\safemath{\vecm}{\bmm}
\safemath{\vecn}{\bmn}
\safemath{\veco}{\bmo}
\safemath{\vecp}{\bmp}
\safemath{\vecq}{\bmq}
\safemath{\vecr}{\bmr}
\safemath{\vecs}{\bms}
\safemath{\vect}{\bmt}
\safemath{\vecu}{\bmu}
\safemath{\vecv}{\bmv}
\safemath{\vecw}{\bmw}
\safemath{\vecx}{\bmx}
\safemath{\vecy}{\bmy}
\safemath{\vecz}{\bmz}

\safemath{\veczero}{\bmzero}
\safemath{\vecone}{\bmone}
\safemath{\vecxi}{\bmxi}
\safemath{\veclambda}{\bmlambda}
\safemath{\vecmu}{\bmmu}
\safemath{\vectheta}{\bmtheta}
\safemath{\vecphi}{\bmphi}
\safemath{\vecdelta}{\bmdelta}

\safemath{\matA}{\bA}
\safemath{\matB}{\bB}
\safemath{\matC}{\bC}
\safemath{\matD}{\bD}
\safemath{\matE}{\bE}
\safemath{\matF}{\bF}
\safemath{\matG}{\bG}
\safemath{\matH}{\bH}
\safemath{\matI}{\bI}
\safemath{\matJ}{\bJ}
\safemath{\matK}{\bK}
\safemath{\matL}{\bL}
\safemath{\matM}{\bM}
\safemath{\matN}{\bN}
\safemath{\matO}{\bO}
\safemath{\matP}{\bP}
\safemath{\matQ}{\bQ}
\safemath{\matR}{\bR}
\safemath{\matS}{\bS}
\safemath{\matT}{\bT}
\safemath{\matU}{\bU}
\safemath{\matV}{\bV}
\safemath{\matW}{\bW}
\safemath{\matX}{\bX}
\safemath{\matY}{\bY}
\safemath{\matZ}{\bZ}
\safemath{\matzero}{\bmzero}

\safemath{\matDelta}{\bDelta}
\safemath{\matLambda}{\bLambda}
\safemath{\matPhi}{\bPhi}
\safemath{\matSigma}{\bSigma}
\safemath{\matOmega}{\bOmega}
\safemath{\matTheta}{\bTheta}

\safemath{\matidentity}{\matI}
\safemath{\matone}{\matO}

\safemath{\rnda}{A}
\safemath{\rndb}{B}
\safemath{\rndc}{C}
\safemath{\rndd}{D}
\safemath{\rnde}{E}
\safemath{\rndf}{F}
\safemath{\rndg}{G}
\safemath{\rndh}{H}
\safemath{\rndi}{I}
\safemath{\rndj}{J}
\safemath{\rndk}{K}
\safemath{\rndl}{L}
\safemath{\rndm}{M}
\safemath{\rndn}{N}
\safemath{\rndo}{O}
\safemath{\rndp}{P}
\safemath{\rndq}{Q}
\safemath{\rndr}{R}
\safemath{\rnds}{S}
\safemath{\rndt}{T}
\safemath{\rndu}{U}
\safemath{\rndv}{V}
\safemath{\rndw}{W}
\safemath{\rndx}{X}
\safemath{\rndy}{Y}
\safemath{\rndz}{Z}

\safemath{\rveca}{\bimA}
\safemath{\rvecb}{\bimB}
\safemath{\rvecc}{\bimC}
\safemath{\rvecd}{\bimD}
\safemath{\rvece}{\bimE}
\safemath{\rvecf}{\bimF}
\safemath{\rvecg}{\bimG}
\safemath{\rvech}{\bimH}
\safemath{\rveci}{\bimI}
\safemath{\rvecj}{\bimJ}
\safemath{\rveck}{\bimK}
\safemath{\rvecl}{\bimL}
\safemath{\rvecm}{\bimM}
\safemath{\rvecn}{\bimN}
\safemath{\rveco}{\bomO}
\safemath{\rvecp}{\bimP}
\safemath{\rvecq}{\bimQ}
\safemath{\rvecr}{\bimR}
\safemath{\rvecs}{\bimS}
\safemath{\rvect}{\bimT}
\safemath{\rvecu}{\bimU}
\safemath{\rvecv}{\bimV}
\safemath{\rvecw}{\bimW}
\safemath{\rvecx}{\bimX}
\safemath{\rvecy}{\bimY}
\safemath{\rvecz}{\bimZ}

\safemath{\rvecxi}{\bmxi}
\safemath{\rveclambda}{\bmlambda}
\safemath{\rvecmu}{\bmmu}
\safemath{\rvectheta}{\bmtheta}
\safemath{\rvecphi}{\bmphi}

\safemath{\rmatA}{\bimA}
\safemath{\rmatB}{\bimB}
\safemath{\rmatC}{\bimC}
\safemath{\rmatD}{\bimD}
\safemath{\rmatE}{\bimE}
\safemath{\rmatF}{\bimF}
\safemath{\rmatG}{\bimG}
\safemath{\rmatH}{\bimH}
\safemath{\rmatI}{\bimI}
\safemath{\rmatJ}{\bimJ}
\safemath{\rmatK}{\bimK}
\safemath{\rmatL}{\bimL}
\safemath{\rmatM}{\bimM}
\safemath{\rmatN}{\bimN}
\safemath{\rmatO}{\bimO}
\safemath{\rmatP}{\bimP}
\safemath{\rmatQ}{\bimQ}
\safemath{\rmatR}{\bimR}
\safemath{\rmatS}{\bimS}
\safemath{\rmatT}{\bimT}
\safemath{\rmatU}{\bimU}
\safemath{\rmatV}{\bimV}
\safemath{\rmatW}{\bimW}
\safemath{\rmatX}{\bimX}
\safemath{\rmatY}{\bimY}
\safemath{\rmatZ}{\bimZ}

\safemath{\rmatDelta}{\bimDelta}
\safemath{\rmatLambda}{\bimLambda}
\safemath{\rmatPhi}{\bimPhi}
\safemath{\rmatSigma}{\bimSigma}
\safemath{\rmatOmega}{\bimOmega}
\safemath{\rmatTheta}{\bimTheta}

\usepackage{amssymb}
\usepackage{amsfonts}
\usepackage{mathrsfs}
\usepackage{xspace}
\usepackage{bm}
\usepackage{fancyref}
\usepackage{textcomp}

\usepackage{multirow}
\usepackage{stmaryrd}

\newenvironment{textbmatrix}{	\setlength{\arraycolsep}{2.5pt}%
								\big[\begin{matrix}}{\end{matrix}\big]%
								\raisebox{0.08ex}{\vphantom{M}}}

\def\be{\begin{equation}}
\def\ee{\end{equation}}
\def\een{\nonumber \end{equation}}
\def\mat{\begin{bmatrix}}
\def\emat{\end{bmatrix}}
\def\btm{\begin{textbmatrix}}
\def\etm{\end{textbmatrix}}

\def\ba#1\ea{\begin{align}#1\end{align}}
\def\bas#1\eas{\begin{align*}#1\end{align*}}
\def\bs#1\es{\begin{split}#1\end{split}}
\def\bg#1\eg{\begin{gather}#1\end{gather}}
\def\bml#1\eml{\begin{multline}#1\end{multline}}
\def\bi#1\ei{\begin{itemize}#1\end{itemize}}

\DeclareMathOperator*{\argmax}{arg\;max}		

\safemath{\dirac}{\delta}					
\safemath{\krond}{\dirac}					

\safemath{\upto}{\uparrow}
\safemath{\downto}{\downarrow}
\safemath{\iu}{j}							
\safemath{\ev}{\lambda}						
\safemath{\hilseqspace}{l^{2}}				
\newcommand{\banachfunspace}[1]{\setL^{#1}}	
\safemath{\hilfunspace}{\banachfunspace{2}}	

\safemath{\SNR}{\textit{SNR}} 				
\safemath{\PAR}{\textit{PAR}} 				
\safemath{\No}{N_0}							
\safemath{\Es}{E_s}							
\safemath{\Eb}{E_b}							
\safemath{\EbNo}{\frac{\Eb}{\No}}
\safemath{\EsNo}{\frac{\Es}{\No}}

\DeclareMathOperator{\CHop}{\ensuremath{\opH}} 
\safemath{\tvir}{\rndh_{\CHop}}				
\safemath{\tvtf}{\rndl_{\CHop}}				
\safemath{\spf}{\rnds_{\CHop}}				
\safemath{\bff}{H_{\CHop}}					

\safemath{\ircf}{r_{h}}						
\safemath{\tftvcf}{r_{s}}					
\safemath{\tfcf}{r_{l}}						
\safemath{\bfcf}{r_{H}}						

\safemath{\tcorr}{c_h}						
\safemath{\scf}{c_{s}}						
\safemath{\tfcorr}{c_{l}}					
\safemath{\fcorr}{c_{H}}						

\safemath{\mi}{I}							
\safemath{\capacity}{C}						

\safemath{\normal}{\mathcal{N}}			
\safemath{\jpg}{\mathcal{CN}}			
\safemath{\mchain}{\leftrightarrow}		

\safemath{\dB}{\,\mathrm{dB}}
\safemath{\dBm}{\,\mathrm{dBm}}
\safemath{\Hz}{\,\mathrm{Hz}}
\safemath{\kHz}{\,\mathrm{kHz}}
\safemath{\MHz}{\,\mathrm{MHz}}
\safemath{\GHz}{\,\mathrm{GHz}}
\safemath{\s}{\,\mathrm{s}}
\safemath{\ms}{\,\mathrm{ms}}
\safemath{\mus}{\,\mathrm{\text{\textmu}s}}
\safemath{\ns}{\,\mathrm{ns}}
\safemath{\ps}{\,\mathrm{ps}}
\safemath{\meter}{\,\mathrm{m}}
\safemath{\mm}{\,\mathrm{mm}}
\safemath{\cm}{\,\mathrm{cm}}
\safemath{\m}{\,\mathrm{m}}
\safemath{\W}{\,\mathrm{W}}
\safemath{\mW}{\, \mathrm{mW}}
\safemath{\J}{\,\mathrm{J}}
\safemath{\K}{\,\mathrm{K}}
\safemath{\bit}{\,\mathrm{bit}}
\safemath{\nat}{\,\mathrm{nat}}

\safemath{\define}{\triangleq}			

\safemath{\equivalent}{\sim}
\safemath{\distas}{\sim}					
\safemath{\sdiff}{\Delta}				

\safemath{\reals}{\mathbb{R}}
\safemath{\positivereals}{\reals_{+}}
\safemath{\integers}{\mathbb{Z}}
\safemath{\posint}{\integers_{+}}
\safemath{\naturals}{\mathbb{N}}
\safemath{\posnaturals}{\naturals_{+}}
\safemath{\complexset}{\mathbb{C}}
\safemath{\rationals}{\mathbb{Q}}

\newcommand*{\fancyrefapplabelprefix}{app}		
\newcommand*{\fancyrefthmlabelprefix}{thm}		
\newcommand*{\fancyreflemlabelprefix}{lem}		
\newcommand*{\fancyrefcorlabelprefix}{cor}		
\newcommand*{\fancyrefdeflabelprefix}{def}		
\newcommand*{\fancyrefproplabelprefix}{prop}		
\newcommand*{\fancyrefexmpllabelprefix}{exmpl}
\newcommand*{\fancyrefalglabelprefix}{alg}		
\newcommand*{\fancyreftbllabelprefix}{tbl}		

\frefformat{vario}{\fancyrefseclabelprefix}{Sec.~#1}
\frefformat{vario}{\fancyrefthmlabelprefix}{Thm.~#1}
\frefformat{vario}{\fancyreftbllabelprefix}{Tbl.~#1}
\frefformat{vario}{\fancyreflemlabelprefix}{Lem.~#1}
\frefformat{vario}{\fancyrefcorlabelprefix}{Cor.~#1}
\frefformat{vario}{\fancyrefdeflabelprefix}{Def.~#1}
\frefformat{vario}{\fancyreffiglabelprefix}{Fig.~#1}
\frefformat{vario}{\fancyrefapplabelprefix}{App.~#1}
\frefformat{vario}{\fancyrefeqlabelprefix}{(#1)}
\frefformat{vario}{\fancyrefproplabelprefix}{Prop.~#1}
\frefformat{vario}{\fancyrefexmpllabelprefix}{Ex.~#1}
\frefformat{vario}{\fancyrefalglabelprefix}{Alg.~#1}

\safemath{\dictab}{[\,\dicta\,\,\dictb\,]}

\safemath{\ysig}{\bmy}
\safemath{\ysighat}{\hat{\ysig}}
\safemath{\ysigdim}{M}
\safemath{\xsig}{\bmx}
\safemath{\xsigdim}{N}
\safemath{\nx}{n_x}
\safemath{\zsig}{\bmz}
\safemath{\zsigdim}{\ysigdim}
\safemath{\rsig}{\bmr}
\safemath{\Adict}{\bA}
\safemath{\Adicttilde}{\widetilde{\Adict}}
\safemath{\Adictdim}{\outputdim\times\xsigdim}
\safemath{\avec}{\bma}
\safemath{\avectilde}{\tilde{\avec}}
\safemath{\Bdict}{\bB}
\safemath{\Bdicttilde}{\widetilde{\Bdict}}
\safemath{\Cdict}{\bC}
\safemath{\cvec}{\bmc}
\safemath{\Ddict}{\bD}
\safemath{\Ddictdim}{\ysigdim\times\xsigdim}
\safemath{\dvec}{\bmd}
\safemath{\Ddicttilde}{\widetilde{\bD}}
\safemath{\Bonb}{\bB}
\safemath{\bvec}{\bmb}
\safemath{\Bonbdim}{\ysigdim\times\ysigdim}
\safemath{\noise}{\bmn}
\safemath{\noisedim}{\ysigim}
\safemath{\err}{\bme}
\safemath{\errdim}{\ysigdim}
\safemath{\errset}{\setE}
\safemath{\nerr}{n_e}
\safemath{\delop}{\bP_\errset}
\safemath{\delopc}{\bP_{{\errset}^c}}

\safemath{\cplxi}{\imath}
\safemath{\cplxj}{\jmath}

\safemath{\dict}{\matD}
\safemath{\inputdim}{N}		
\safemath{\outputdim}{M}		
\safemath{\sparsity}{S}	
\safemath{\inputdimA}{{N_a}}	
\safemath{\inputdimB}{{N_b}}	
\safemath{\elemA}{{n_a}}	
\safemath{\elemB}{{n_b}}	
\safemath{\resA}{\matR_a}	
\safemath{\resB}{\matR_b}	
\safemath{\subD}{\matS} 
\safemath{\subA}{\matS_a} 
\safemath{\subB}{\matS_b} 
\safemath{\dicta}{\matA} 	
\safemath{\dictb}{\matB} 	
\safemath{\hollowS}{H}
\safemath{\hollowA}{H_a}
\safemath{\hollowB}{H_b}
\safemath{\cross}{Z}
\safemath{\coh}{\mu_d}			
\safemath{\coha}{\mu_a}			
\safemath{\cohb}{\mu_b}			
\safemath{\mubs}{\nu}	
\safemath{\cohm}{\mu_m} 
\safemath{\dictset}{\setD}	
\safemath{\dictsetp}{\dictset(\coh,\coha,\cohb)}	
\safemath{\dictsetgen}{\dictset_\text{gen}}
\safemath{\dictsetgenp}{\dictsetgen(\coh)}
\safemath{\dictsetonb}{\dictset_\text{onb}}
\safemath{\dictsetonbp}{\dictsetonb(\coh)}

\safemath{\leftside}{U}
\safemath{\rightsideA}{R_a}
\safemath{\rightsideB}{R_b}

\safemath{\indexS}{\setI_S} 

\safemath{\na}{n_a}			
\safemath{\nb}{n_b}			
\safemath{\coeffa}{p_i}	
\safemath{\coeffb}{q_j}	
\safemath{\seta}{\setP}		
\safemath{\setb}{\setQ}     
\safemath{\setw}{\setW}	
\safemath{\setz}{\setZ}	
\safemath{\cola}{\veca}		
\safemath{\colb}{\vecb}		
\safemath{\cold}{\vecd}		
\safemath{\inputvec}{\vecx} 	
\safemath{\error}{\vece}	
\safemath{\noiseout}{\vecz} 	
\safemath{\inputvecel}{x}
\safemath{\inputveca}{\vecx_a}
\safemath{\inputvecb}{\vecx_b}
\safemath{\outputvec}{\vecy}	
\safemath{\lambdamin}{\lambda_{\mathrm{min}}}

\safemath{\elltwo}{\ell_2}
\safemath{\ellone}{\ell_1}
\safemath{\ellzero}{\ell_0}
\safemath{\ellinf}{\ell_\infty}
\safemath{\ellinftilde}{\ell_{\widetilde\infty}}
\safemath{\licard}{Z(\coh,\coha,\cohb)}
\safemath{\xsol}{\hat{x}}
\safemath{\xbord}{x_b}		
\safemath{\xstat}{x_s}		
\safemath{\xstatLone}{\tilde{x}_s}
\safemath{\order}{\mathcal{O}} 
\safemath{\scales}{\Theta} 
\safemath{\ones}{\mathbf{1}} 
\safemath{\zeroes}{\mathbf{0}} 
\safemath{\thlone}{\kappa(\coh,\cohb)} 
\safemath{\constoneA}{\delta} 
\safemath{\constoneB}{\epsilon} 
\safemath{\nlarge}{L}				   
\safemath{\sumlarge}{S_\nlarge}
\safemath{\maxlarger}{P_\nlarge}	   
\safemath{\Pzero}{\textrm{P0}}	
\safemath{\Pone}{\textrm{P1}}
\safemath{\vecfir}{\vecw}			 
\safemath{\vecsec}{\vecz}
\safemath{\elvecfir}{w}              
\safemath{\elvecsec}{z}				 
\safemath{\nlargefir}{n}
\safemath{\normout}{\gamma}
\safemath{\auxfun}{h}
\safemath{\supp}{\textrm{supp}}

\safemath{\indexa}{\ell}
\safemath{\indexb}{r}
\safemath{\indexc}{i}
\safemath{\indexd}{j}

\safemath{\project}{P}

\newcommand*{\fancyrefremarklabelprefix}{remark}
\frefformat{vario}{\fancyrefremarklabelprefix}{Remark~#1}

\usepackage{color}

\def\loss{\mathfrak{L}}

\newcommand{\setMs}{\setM_{\textnormal{S}}}
\newcommand{\setMr}{\setM_{\textnormal{R}}}

\IEEEoverridecommandlockouts 
\allowdisplaybreaks 


\title{Channel Charting for Streaming CSI Data
\author{\IEEEauthorblockN{Sueda Taner$^\textnormal{1}$, Maxime Guillaud$^\textnormal{2}$, Olav Tirkkonen$^\textnormal{3}$, and Christoph Studer$^\textnormal{1}$}\\[0.5cm]
\IEEEauthorblockA{\em 
$^\textnormal{1}$Department of Information Technology and Electrical Engineering, ETH Zurich, Switzerland \\
$^\textnormal{2}$Inria \& CITI Laboratory, France \\
$^\textnormal{3}$Department of Communications and Networking, Aalto University, Finland \\
email: taners@iis.ee.ethz.ch, maxime.guillaud@inria.fr, olav.tirkkonen@aalto.fi, and studer@ethz.ch} 
}
\thanks{The authors acknowledge the support of the European CHIST-ERA program through the CHASER (channel charting as a service) project.}}

\begin{document}

\maketitle


\begin{abstract}
Channel charting (CC) applies dimensionality reduction to channel state information (CSI) data at the infrastructure basestation side with the goal of extracting pseudo-position information for each user.
The self-supervised nature of CC enables predictive tasks that depend on user position without requiring any ground-truth position information.  
In this work, we focus on the practically relevant \emph{streaming} CSI data scenario, in which CSI is constantly estimated. 
To deal with storage limitations, we develop a novel streaming CC architecture that maintains a small core CSI dataset from which the channel charts are learned. Curation of the core CSI dataset is achieved using a min-max-similarity criterion.
Numerical validation with measured CSI data demonstrates that our method approaches the accuracy obtained from the complete CSI dataset while using only a fraction of CSI storage and avoiding catastrophic forgetting of old CSI data. 
\end{abstract}



\section{Introduction} 
\label{sec:intro}

Channel charting (CC), proposed in \cite{Studer2018}, applies dimensionality reduction to large sets of estimated channel state information (CSI) data acquired over long time periods. 
For a typical (fixed) wireless infrastructure access point (AP) or basestation (BS) and a fixed scattering environment (buildings etc.), CSI is predominantly influenced by the scattering environment. Thus, low-dimensional embeddings of such CSI datasets of mobile users produced by CC are expected to be topologically similar to certain features of the environment.
This property has been verified experimentally, for instance in \cite{ferrand2021}, where the two-dimensional embedding of the mobile users is topologically similar to the users' position on the ground.
Here, topological similarity means that the obtained representation is accurate up to transformations such as rotating, stretching, and other continuous (but possibly nonlinear) transforms.
Such transformations exist due to the self-supervised nature of CC, which also precludes its use for classical positioning. Nonetheless, CC associates a \emph{pseudo position} to each CSI sample, which can be used for a broad range of location-based tasks---see~\cite{ferrand2023wireless} for an overview of CC and its applications.

Typical applications of CC involve repeatedly performing dimensionality reduction (DR) on large amounts of CSI samples that are continuously collected at high rates (e.g., tens to hundreds times per second) by APs or BSs with limited storage and computing power.
While DR attempts to capture the geometry of the complete distribution, since the beginning of observing CSI samples, storing \emph{all} CSI data will inevitably exhaust the memory available at the AP/BS.
Thus, in practice, only a tiny fraction of the collected CSI data can be stored long-term.
In addition, since successive CSI samples cannot be assumed drawn independently from this distribution, due to the slow evolution of the geometric propagation parameters (e.g., moving users), classical approaches that deal with streaming data, such as time windowing, potentially result in catastrophic forgetting~\cite{kirkpatrick2017}.
Therefore, wireless communication systems that generate a constant stream of new CSI samples require an online method that maintains a fixed dataset size, while still enabling the extraction of high-quality channel charts and avoiding catastrophic forgetting.

\subsection{Contributions} 

Since the CC literature routinely ignores the fact that CSI data is acquired perpetually in a streaming fashion, we introduce a new approach for CC with streaming CSI data and limited memory.
Our method consists of a simple dynamic core dataset selection strategy that updates the set of CSI samples used for learning the channel charts. 
This strategy has the benefit of conceptually separating the distillation algorithm from the DR process, which is performed solely based on the core CSI memory contents; see Fig.~\ref{fig:streaming_pipeline} for an overview of our approach. 
We introduce a core CSI memory curation strategy that dynamically minimizes the maximum similarity across the stored CSI samples.
Results with the measured CSI dataset from~\cite{dichasus2021} demonstrate that the proposed min-max-similarity curation approach avoids catastrophic forgetting and, even with a small core CSI memory, achieves comparable performance as an idealistic baseline that has access to \emph{all} CSI samples.

\subsection{Relevant Prior Art}

In the field of machine learning, the problem of selecting a subset of a dataset while preserving its information content (measured as its potential to be used as a training set yielding a prescribed accuracy) is known as core-set selection \cite{pmlr-v119-mirzasoleiman20a}.
A more general approach, known as dataset condensation (or distillation), consists of synthesizing a new composite, smaller dataset (not necessarily a subset of the full dataset) based on linear combinations of the original samples (see e.g. \cite{zhao2021condensation_gradient_matching}).
Dataset condensation appears to be a valid approach to deal with streaming data (see, e.g., the results in \cite{Wievel_condensed_composite_memory_continual_learning,Sangermano_IJCNN22_sample_condensation_OCL} which leverage dataset condensation through a gradient matching approach for the purpose of continual learning).
However, existing dataset condensation approaches (see \cite{yu2023dataset_distillation} for an overview) appear to be limited to \emph{supervised} learning problems, and their extension to self-supervised DR is not straightforward.
In contrast, our proposed approach is tailored to CC, which is an inherently self-supervised learning problem. 
 

\section{Channel Charting Overview}
\label{sec:background}

We start by introducing the system model and summarizing traditional CC from non-streaming CSI.

\subsection{System Model }
\label{sec:system_model}

We consider a single-input multiple-output (SIMO) wireless communication system, in which one or multiple single-antenna user equipments (UEs) transmit pilots to one or multiple APs/BSs, with a total number of $B$ receive antennas.
We assume that transmission utilizes orthogonal frequency-division multiplexing (OFDM) with $W$ used subcarriers.
For one UE at position $\bmx^{(n)}\in\opR^3$, the CSI vector estimated by the BS at the $w$th subcarrier is $\bmh^{(n)}_w\in\opC^{B}$, where $n$ denotes the sample index.
The CSI matrix associated with the UE at position~$\bmx^{(n)}$ is obtained by concatenating the channel vectors 
from all subcarriers $[\bmh_1^{(n)},\dots,\bmh_W^{(n)}]=\bH^{(n)}\in\opC^{B \times W}$.

\subsection{Channel Charting Basics}

Traditional CC operates in the following two phases~\cite{Studer2018}. 

\subsubsection{Learning the CC Function}
In the first phase, a large CSI dataset from various UE positions is collected at one or many infrastructure basestations (BSs).
To capture the large-scale fading characteristics of the wireless channel~\cite{Studer2018} and to improve resilience against system and hardware impairments\mbox{\cite{ferrand2021, gonultas22twc}}, one first transforms each CSI matrix $\bH^{(n)}$ into a CSI feature~$\vecf^{(n)}\in\reals^{D'}$.
We represent this transform by a feature extraction function $f:\opC^{B\times W} \to \opR^{D'}$, so that $\vecf^{(n)}=f(\bH^{(n)})$.
The total dataset of acquired CSI features is $\{\bmf^{(n)}\}_{n=1}^N$, where~$N$ denotes the total number of acquired CSI samples. 
We note that if CSI is acquired at fast rates and over long periods of time, then storing all $D'\times N$ scalars may be infeasible. 

From the CSI feature dataset, one then learns a CC function $g_{\boldsymbol\theta}:\opR^{D'}\to\opR^{D}$, where $D\ll D'$, so that 
each CSI feature $\vecf^{(n)}$
is mapped to a $D$-dimensional pseudo-position $\hat\vecx^{(n)} = g_{\boldsymbol\theta}(\vecf^{(n)})$.
Here, the vector~$\boldsymbol\theta$ denotes the trainable parameters of the dimensionality reduction function.
Note that the CC function is trained in a self-supervised manner, solely from the CSI feature dataset $\{\bmf^{(n)}\}_{n=1}^N$, which does \emph{not} use any ground-truth UE position information---training typically aims at preserving pairwise relationships (e.g., matching dissimilarities) between feature space and latent space.
The resulting, low-dimensional latent-space representation is the \emph{channel chart}.

\subsubsection{Channel Charting}
In the second phase, the learned CC function $g_{\boldsymbol\theta}$ is then used for channel charting, where one maps (typically in real time) new CSI features to pseudo-positions in the channel chart. The benefit of a channel chart is that it captures pseudo-positions of the transmitting UEs by preserving the local geometry: UEs that are close in space are also close in the channel chart.


\section{Channel Charting with Streaming CSI}

We now present our architecture for CC from streaming CSI and then discuss two core memory curation strategies. 

\subsection{Architecture Overview}

\begin{figure}[tp]
\centering
\includegraphics[width=0.99\columnwidth]{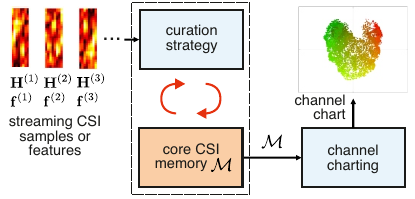}
\caption{Illustration of channel charting for streaming CSI data. A curation strategy maintains a subset of the obtained CSI samples or features in a small core CSI memory $\setM$, which is used to perform channel charting. }
\label{fig:streaming_pipeline}
\end{figure}

In a realistic scenario, APs or BSs continuously estimate CSI matrices at high rates, which prevents them from storing all CSI data over long periods of time.
We address this problem by learning a CC function from streaming CSI under the constraint of limited memory at the BSs.
Concretely, we propose an architecture as depicted in~\fref{fig:streaming_pipeline}. Here, either CSI matrices or features arrive perpetually and in a streaming fashion.
A distillation algorithm, which is responsible for curating a core CSI dataset~$\setM$
that stores no more than~$M$ CSI samples, assesses each newly-arrived CSI sample and either stores it in the core CSI memory (which may require one to replace a datapoint in the core CSI memory) or discards it.\footnote{One could also directly store CSI features in the core CSI memory $\setM$. In this work, we store the CSI matrices as it enables the use of elaborate dissimilarity metrics for channel charting; see~\fref{sec:pipeline} for the details.} 
The CC function $g_{\boldsymbol\theta}$ is then learned from the core CSI dataset $\setM$.

\subsection{Core CSI Memory Curation Strategies}
\label{sec:curationstrategies}

The core CSI memory is then curated by storing only a subset of the streaming CSI data (matrices or features) as follows.
At the beginning, all of the estimated CSI samples are added to the core memory until its capacity $M$ is reached.
After capacity has been reached, we asses each new CSI sample. If we decide to add the new CSI sample to the core CSI memory, then we maintain fixed memory capacity by replacing one existing CSI sample. Otherwise, the memory remains unchanged and the new CSI sample is discarded. 
We now propose two curation strategies that determine whether to add a new CSI sample and which existing CSI sample to discard.

\subsubsection{Random Subset (RandoS)}

The first (and simplest) curation strategy that would come to one's mind is to store a random subset of the streaming CSI.
Concretely, after core memory capacity has been reached, this curation strategy randomly decides whether to update the core memory by adding a new CSI sample $\bH^{(n)}$ with probability $p_{\textnormal{update}}\in[0,1]$. 
If one decides to add the new CSI sample, then one chooses to randomly replace the $r$th element $\bH^{(r)}\in\setM$ according to the probability mass function (PMF) $p_r\in[0,1]$ with  $\sum_{r} p_r = 1$.
One can choose $p_{\textnormal{update}}$ and the PMF $p_r$, $r=1,\ldots,M$, according to some preference (e.g., updating the core memory more frequently or replacing more recent CSI samples).

\subsubsection{Similarity-Based Subset (SimS)}

The second curation strategy takes into account the CC application and attempts to store a subset of the streaming CSI that minimizes the maximum similarity in the core memory.
The intuition behind this strategy is to avoid redundancy by maintaining a core CSI memory $\setM$ with maximally dissimilar CSI samples. 
The implementation details of this curation strategy are given in~\fref{alg:sims} and the method proceeds as follows. 
For each new arriving CSI sample $\bH^{(n)}$, after memory capacity $M$ is reached, we first calculate the similarity between the new CSI sample and all of the CSI samples in the core memory (line~6).
By denoting the $i$th element in the core CSI memory $\setM$ by $\bM_i$,
we find the pair from $\setM$ that exhibits maximum similarity $\{\bM_k,\bM_\ell\}$; this set contains the two candidates of samples that should be replaced (line~7).
If the maximum of the similarities between the new CSI sample and the CSI samples in the core memory is smaller than the similarity between $\bM_k$ and $\bM_\ell$ (line~8), then, 
with probability~$p$, we replace $\bM_k$ with the new sample~$\bH^{(n)}$; otherwise, we replace~$\bM_\ell$ (lines~9 and~10).
In our implementation, we measure the pairwise CSI similarity using the absolute cosine similarity between CSI features: 
\newcommand{\simfunc}{\mathrm{sim}}
\begin{align}\label{eq:cossim}
\simfunc(\bH^{(i)}, \bH^{(j)}) = \frac{ |({\vecf^{(i)})^H \vecf^{(j)} } |}{ {\|\vecf^{(i)}\| \|\vecf^{(j)}\|} }.
\end{align}
Here, $\vecf^{(i)} = f(\bH^{(i)})$, the superscript $^H$ is the conjugate transpose, and $\|\cdot\|$ is the Euclidean norm. 

\begin{algorithm}[tp]
\begin{algorithmic}[1]
\caption{Similarity-based Subset (SimS) Curation Strategy}
\label{alg:sims}
\State {\bf initialize} $\setM = \varnothing$ and $n=1$
\For{each newly arrived CSI sample $\bH^{(n)}$}
\If{$n\leq M$}
\State $\setM \gets \setM \cup \bH^{(n)}$
\Else{}  
\State $s \gets \max_{{\bM_i}\in\setM} \simfunc(\bH^{(n)}, \bM_i)$ \label{eq:s_n}
\State $\{k,\ell\} \gets \underset{1\leq i\neq j \leq M}{\argmax} \ {\simfunc(\bM_i, \bM_j)}$
\If{$s < \simfunc(\bM_k, \bM_\ell)$} 
\State with probability $p$, $r \gets k$; otherwise, $r \gets \ell$
\State $\setM \gets (\setM \setminus \bM_r) \cup \bH^{(n)}$
\EndIf
\EndIf
\State $n \gets n+1$
\EndFor
\end{algorithmic}
\end{algorithm}


\section{Experimental Setup}

We now describe our experimental setup, which we use in \fref{sec:results} to assess the efficacy of the proposed methods. 

\subsection{Dataset Description}
\label{sec:scenario}

We use the measured CSI from the DICHASUS dataset~\cite{dichasus2021,dataset-dichasus-cf0x}.
These measurements are for an indoor distributed SIMO communication system, in which a moving robot transmits pilots to four APs with eight antennas each.
The APs are synchronized in frequency, time, and phase via over-the-air synchronization, and the transmitting robot uses an omnidirectional antenna.
The system uses OFDM with a total number of $W=1024$ subcarriers over a bandwidth of $50$\,MHz, centered at a carrier frequency of $1.272$\,GHz.
The dataset in~\cite{dataset-dichasus-cf0x} comprises CSI from multiple trajectories of the robot over the same area.

\begin{figure}
    \centering
    \includegraphics[height=5cm]{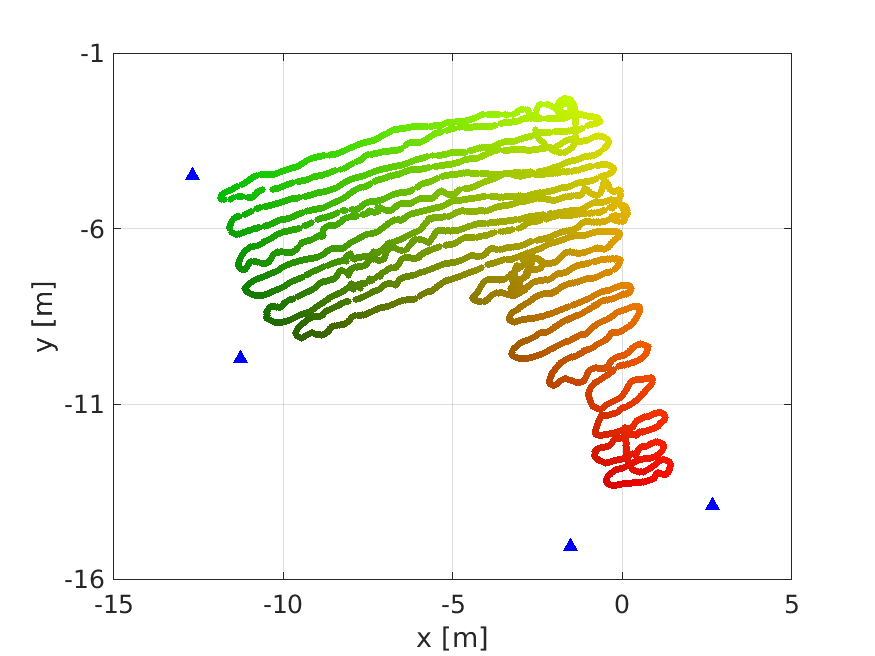}
    \caption{Ground-truth UE positions (green-to-red gradient-colored area) of the streaming CSI and AP positions (blue triangles). Only a subset of the CSI samples stemming from these locations will be stored in the core memory $\setM$. }
    \label{fig:streaming_gt_pos}
\end{figure}

To imitate a streaming CSI scenario, we use only the first trajectory from~\cite{dataset-dichasus-cf0x}. This trajectory yields a total number of $N=17\,516$ sequentially-arriving CSI samples, where the robot follows a path from the top-left corner of the area to the bottom-right\footnote{For easier interpretation of the generated results, we exclude the last $1\,000$ CSI samples where the robot moves back towards the start.}; see~\fref{fig:streaming_gt_pos} for the robot's trajectory. 

\subsection{Channel Charting Pipeline}    
\label{sec:pipeline}

The specifics of our channel charting pipeline are as follows. 
We adopt the CSI feature extraction function from~\cite{lundpaper,lei19siamese, taner2023globecom}, where we first apply an inverse discrete Fourier transform over the $W$  subcarriers to transform the CSI into the delay domain.
We denote the delay domain CSI vector for the $n$th UE position at delay tap $\tau$ with $\underline{\vech}_\tau^{(n)}\in\opC^B$.
We only keep the first $C$ columns of the delay-domain CSI, $\{\underline{\vech}_\tau^{(n)}\}_{\tau=1}^C$, since the first few taps usually contain most of the received power. 
We vectorize this truncated-delay domain CSI matrix and take entrywise absolute values.
Finally, we scale the resulting vector to unit Euclidean norm.
The resulting CSI feature vectors $\vecf^{(n)}\in\opR^{BC}$ will be the input of the CC function.

The literature describes a variety of neural-network-based channel charting functions building on autoencoders~\cite{Studer2018,penghzipaper}, networks trained with a triplet loss~\cite{ferrand2021,yassine22,rappaport2021,taner2023globecom}, or Siamese neural networks \cite{lei19siamese,stahlke23,stephan2023adp}. 
In what follows, we focus on Siamese neural networks that match the distances between the positions in the channel chart to pairwise dissimilarities computed from CSI. 
This property is expressed by the following loss: 
\begin{align}
\loss(\boldsymbol\theta) = \sum_{i=1}^{N-1} \sum_{j=i+1}^{N} \big( d_{i,j}  - \|g_{\boldsymbol\theta} (\vecf^{(i)}) - g_{\boldsymbol\theta} (\vecf^{(j)}) \| \big)^2. \label{eq:loss_siamese}
\end{align}
Here, $d_{i,j}$ denotes the dissimilarity between $\bH^{(i)}$ and $\bH^{(j)}$ used for channel charting.
The measure of dissimilarity we use is calculated in two steps:
First, we compute the angle-delay profile (ADP)-based metric from~\cite{stephan2023adp}, which is given by 
\begin{align}
\tilde{d}_{i,j} = \sum_{\tau=1}^{C}  \bigg( 1 - \frac{ |(\underline{\vech}_\tau^{(i)})^H \underline{\vech}_\tau^{(j)} |^2}{\|\underline{\vech}_\tau^{(i)} \|^2  \|\underline{\vech}_\tau^{(j)}\|^2}  \bigg)  .
\label{eq:d_adp}
\end{align}  
Intuitively, this metric calculates the squared cosine similarity between two delay domain CSI vectors per-tap, and sums over the taps\footnote{In case of a distributed scenario with multiple APs, one would use the CSI vectors per AP and sum over the APs.}. 
Second, we compute geodesic dissimilarities as in~\cite{stahlke23,stephan2023adp}. To this end, we form a $K$-nearest neighbor graph of every sample according to its ADP-based dissimilarities with all other CSI samples.
We then apply Dijkstra's algorithm~\cite{dijkstra1959note} on this graph to determine all shortest paths. 
Finally, the geodesic dissimilarity $d_{i,j}$ is given by the length of the shortest path between samples $i$ and $j$.

The CC function $g_{\boldsymbol\theta}$ is implemented using the neural network architecture from~\cite{ferrand2021}. 
We train a six-layer fully-connected neural network with the following numbers of activations per layer: $\{256, 128, 64, 32, 16, 2\}$. All layers except the last one use ReLU activations; the last one uses linear activations.
We use Glorot's method \cite{glorot} for weight initialization, and we utilize Adam~\cite{kingma2014adam} to train the channel charting function~$g_{\boldsymbol\theta}$.

\subsection{Core-Memory Size and Algorithm Parameters}

We assume that the core memory capacity is $M=1\,000$ and we store a subset of the streaming CSI using the two curation strategies proposed in~\fref{sec:curationstrategies}.
We refer to the core CSI memories obtained with RandoS and SimS by $\setMr$ and $\setMs$, respectively, to avoid ambiguity when we discuss one of them specifically.
For RandoS, we set the probability $p_{\textnormal{update}}=0.5$ and $p_r=1/M$ for all $\bH^{(r)}\in\setMr$. For SimS, we set $p=0.5$.

\subsection{Performance Metrics}
\label{sec:perf_metrics}

We assess the effectiveness of the proposed methods using four standard metrics from the CC literature; more details can be found in~\cite{altous22asilomar}.
(i) Trustworthiness (TW) penalizes false neighborhood relationships in the channel chart, i.e., points that are neighbors in the channel chart but not in real world coordinates. 
(ii) Continuity (CT) quantifies preservation of real-world coordinate neighborhood relationships in the channel chart.
(iii) Kruskal stress (KS) characterizes the mismatch between pairwise distances in real-world coordinates and those in the channel chart. 
(iv) Rajski distance (RD) measures the discrepancy between mutual information and joint entropy of the distribution of pairwise distances in the real-world coordinates and channel chart.
All metrics range from $0$ to $1$. The optimal value for TW and CT is $1$ (large is good); the optimal value for KS and RD is $0$ (small is good).

\section{Results}
\label{sec:results}

We are finally ready to demonstrate the efficacy of our approach to streaming CSI data.

\subsection{Evolution of Core CSI Memory Contents}
\label{sec:construct_memory}

Figures~\ref{fig:memory_evolution_randos} and \ref{fig:memory_evolution_sims} show the ground-truth positions corresponding to the CSI in the core memory after $n=6\,000$, $n=12\,000$, and $n=17\,516$ arrived CSI samples.
The color scheme represents the maximum cosine similarity from \fref{eq:cossim} of each sample with all others in the core memory, i.e., the color of UE position $\vecx^{(n)}$ is determined by $\max_{i\in\setM}\simfunc(\bH^{(n)}, \bH^{(i)})$. 

In~\fref{fig:memory_evolution_randos}, we see that the core memory $\setMr$ formed by RandoS experiences catastrophic forgetting---the memory stores mostly recent (and thus, similar) CSI data and old CSI samples are lost.
In~\fref{fig:memory_evolution_sims}, we see that the core memory $\setMs$ formed by SimS avoids catastrophic forgetting and contains CSI samples with dissimilar cosine similarities. 
We note that the maximum similarity values are generally lower for SimS in~\fref{fig:memory_evolution_sims}(c) than for the earlier time instants in (a) and (b), which is to be expected as the algorithm gradually reduces the maximum cosine similarity within the core memory over time.
However, the maximum similarities of some samples seem to be larger in~\fref{fig:memory_evolution_sims}\,(b) than in (a); this shows that, although the SimS curation strategy aims at reducing the maximum similarity among all samples, the per-sample similarity may increase.

\newcommand{\figheight}{4.5cm}

\begin{figure*}[tp]
    \centering
    \subfigure[$\setMr$ at $n=6\,000$]  
	{
    \includegraphics[height=\figheight]{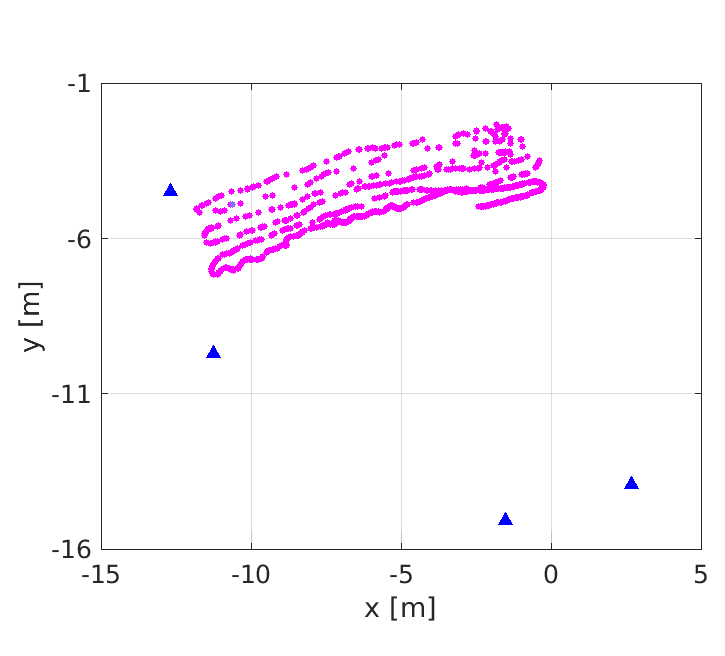}
    }\label{fig:randos_update6k}
    \subfigure[$\setMr$ at $n=12\,000$]  
	{
    \includegraphics[height=\figheight]{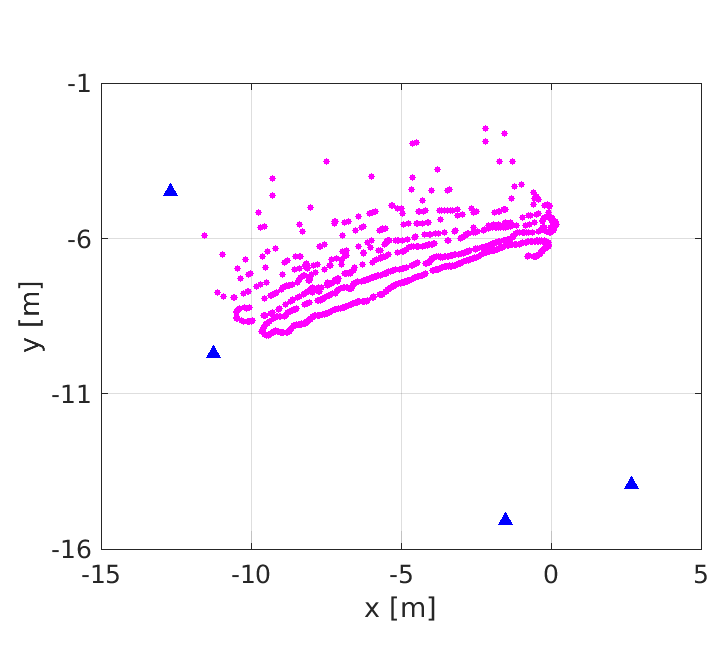}
    }\label{fig:randos_update12k}
    \subfigure[$\setMr$ at $n=N=17\,516$]  
	{
    \includegraphics[height=\figheight]{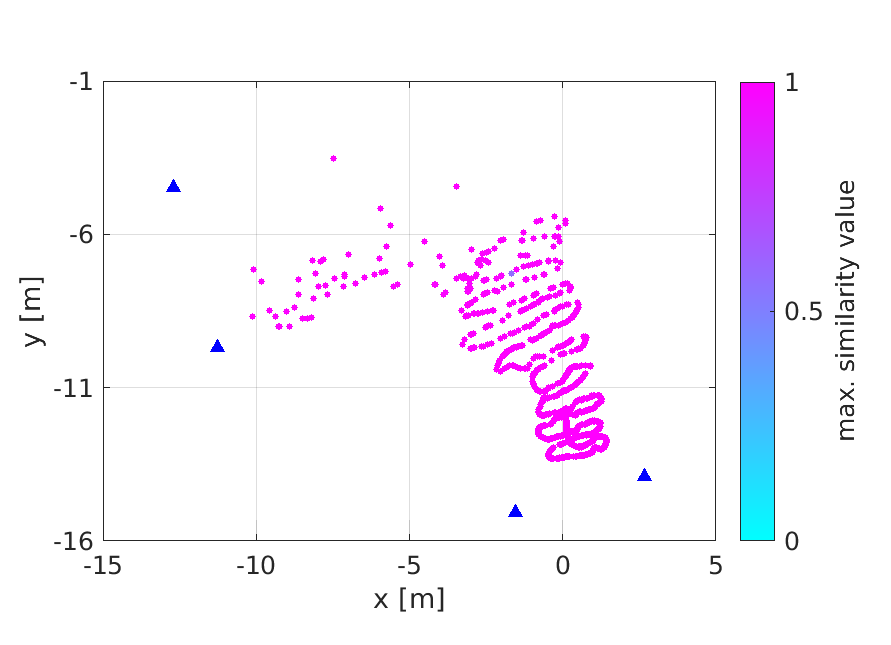}
    }\label{fig:randos_updatelast}
    \caption{Core memory evolution for RandoS. We show ground truth UE positions (cyan-to-pink gradient-colored area) corresponding to the CSI samples stored in the core memory $\setMr$ after $n=6\,000$, $n=12\,000$, and $n=17\,516$ arrived samples. The colors are determined by the maximum similarity of each sample with the other samples in the memory; triangles designate  AP positions.} 
    \vspace{-0.5cm}
    \label{fig:memory_evolution_randos}   
\end{figure*}


\begin{figure*}[tp]
    \centering
    \subfigure[$\setMs$ at $n=6\,000$]  
	{
    \includegraphics[height=\figheight]{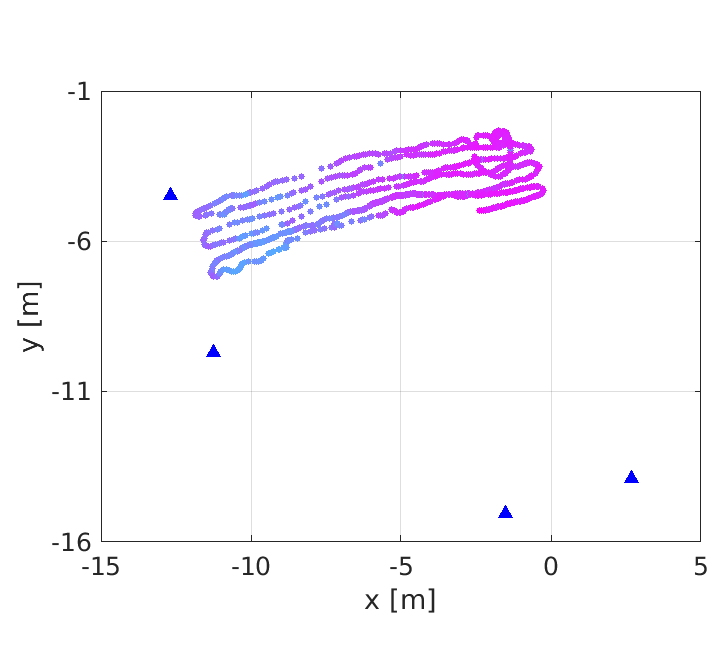}
    }\label{fig:sims_update6k}
    \subfigure[$\setMs$ at $n=12\,000$]  
	{
    \includegraphics[height=\figheight]{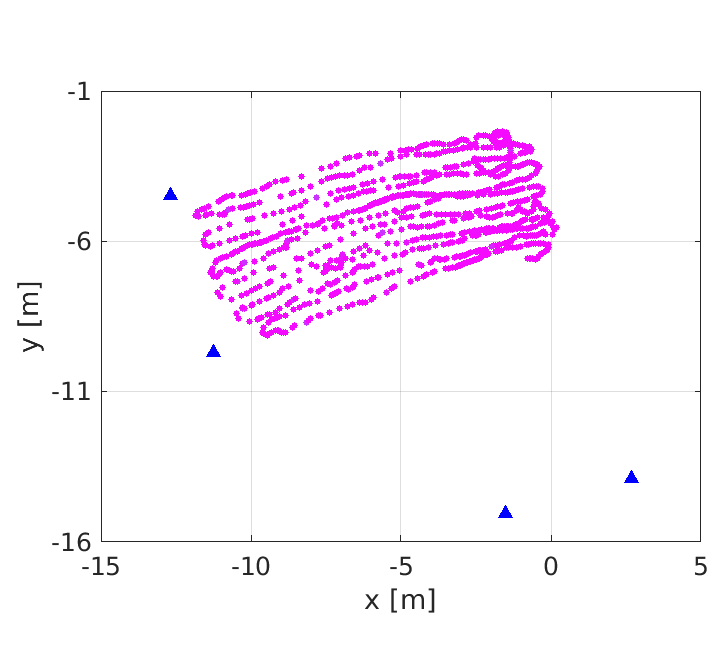}
    }\label{fig:sims_update12k}
    \subfigure[$\setMs$ at $n=N=17\,516$]  
	{
    \includegraphics[height=\figheight]{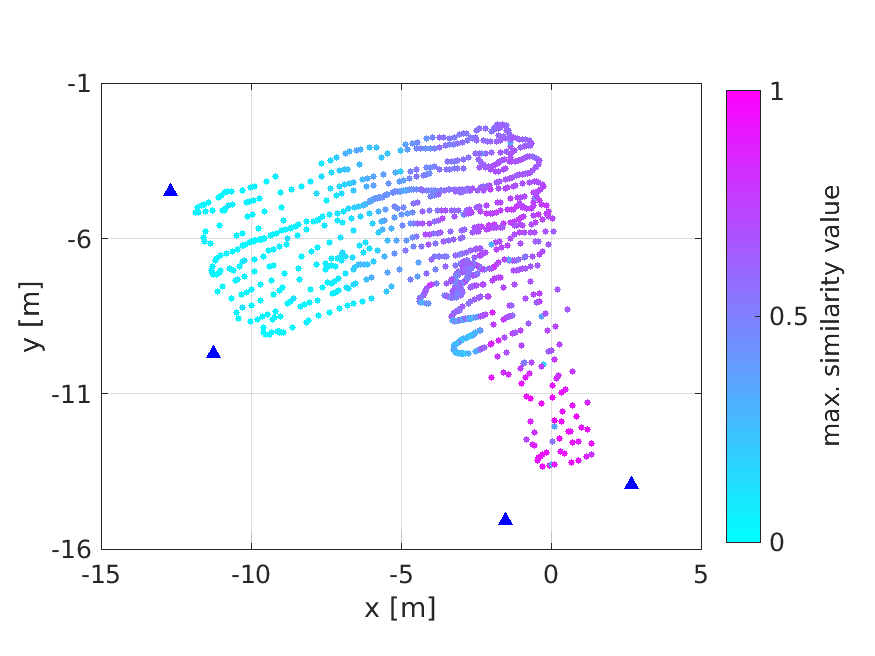}
    }\label{fig:sims_updatelast}
    \caption{Core memory evolution for SimS. We show ground truth UE positions (cyan-to-pink gradient-colored area) corresponding to the CSI samples stored in core memory $\setMs$ after $n=6\,000$, $n=12\,000$, and $n=17\,516$ arrived samples. The colors are determined by the maximum cosine similarity of each CSI sample with all the other samples in the core memory; triangles designate  AP positions.} 
    \label{fig:memory_evolution_sims}   
\end{figure*}

\newcommand{\figsize}{0.43}

\begin{figure*}[tp]
    \centering
    \subfigure[Ground truth positions]  
	{
    \includegraphics[width=\figsize\textwidth]{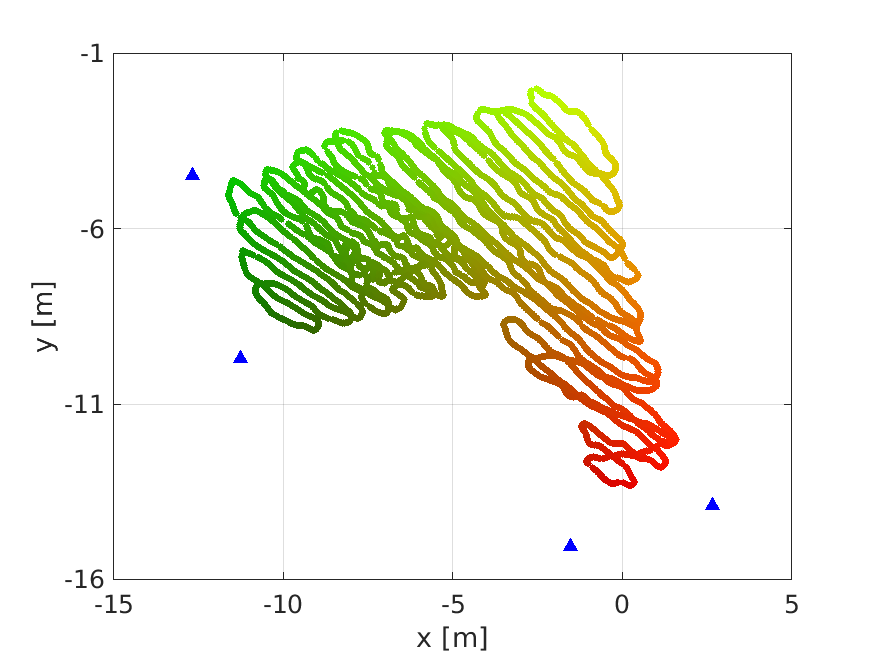}
    }\label{fig:test_gt}
    \hspace{1.5cm}
    \subfigure[Channel chart using RandoS with $\setMr$]  
	{
    \includegraphics[width=\figsize\textwidth]{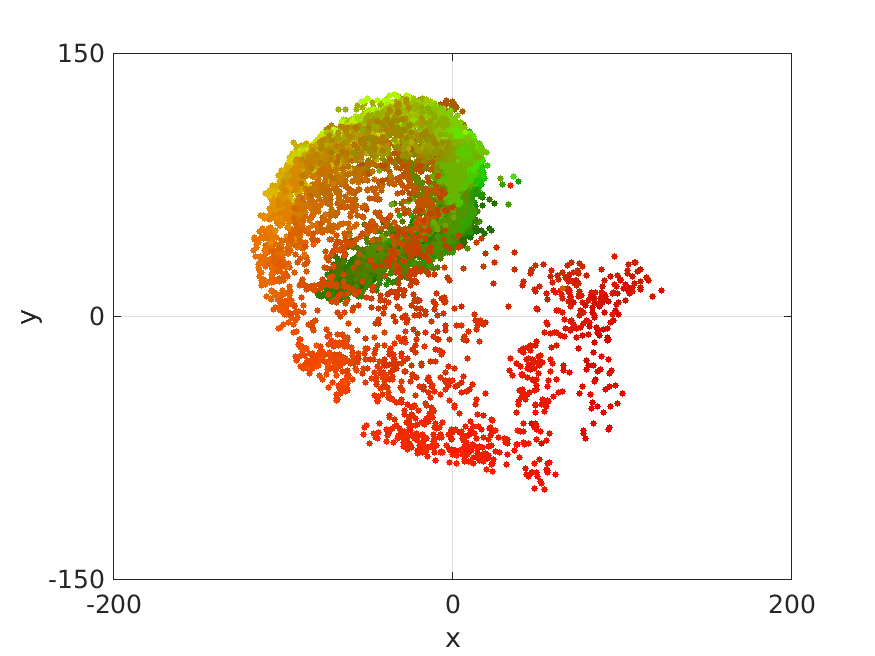}
    }\label{fig:test_cc_randos}   
    \subfigure[Channel chart using SimS with $\setMs$]  
	{
    \includegraphics[width=\figsize\textwidth]{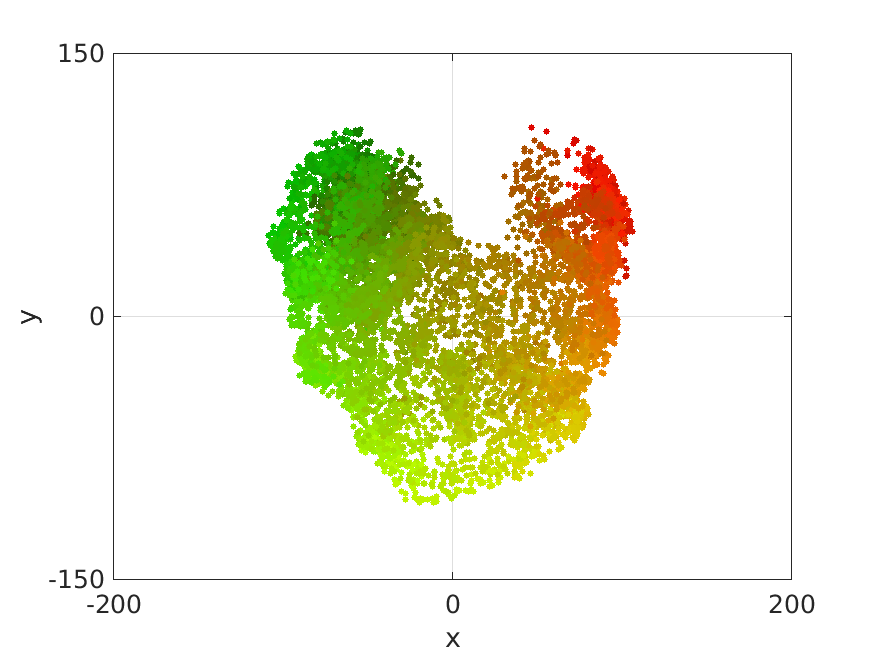}
    }\label{fig:test_cc_sims}
    \hspace{1.5cm}
    \subfigure[Channel chart using all CSI samples]  
	{
    \includegraphics[width=\figsize\textwidth]{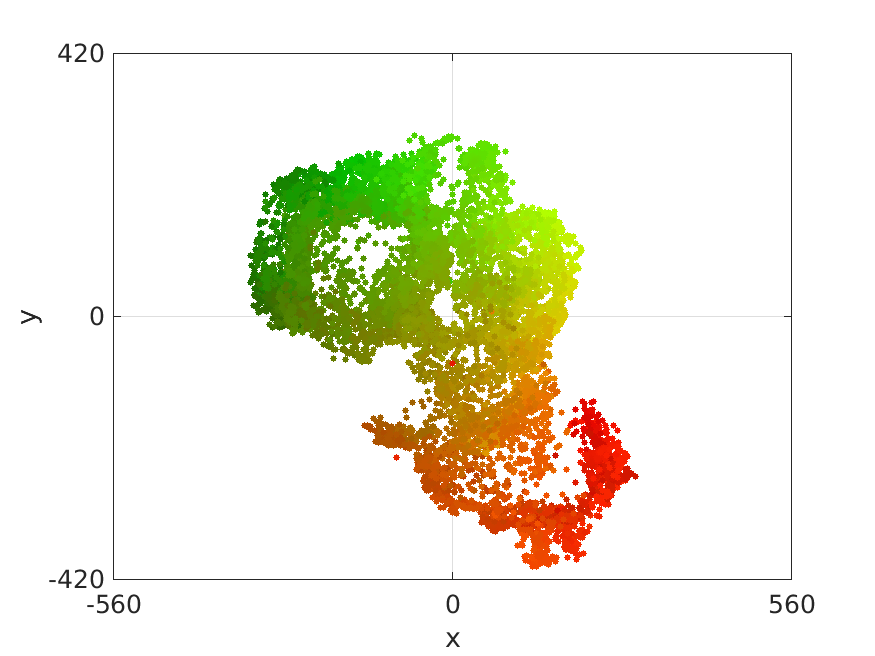}
    }\label{fig:test_cc_all}
    
    \caption{Ground truth positions (a) and channel charts (b-d) for the test set. In~(b) and (c), the channel charting-performing neural networks are trained from the core CSI memories $\setMr$ and $\setMs$, respectively. In (d), all $N= 17\,516$ samples of streaming CSI are used for CC function training.}
    \label{fig:test_charts}   
\end{figure*}

\begin{table}
\centering
\caption{Channel charting performance comparison.}
\label{tbl:result_table}		 
\begin{tabular}{@{}lccccc@{}}
    \toprule
    &  & \multicolumn{4}{c}{Latent space quality metrics} \\
    \cmidrule(lr){3-6}   
    Method & Figure & TW$\,\uparrow$ & CT$\,\uparrow$ & KS$\,\downarrow$ & RD$\,\downarrow$  \\
    \midrule
    RandoS & {5\,(b)}& 0.834 & 0.884 & 0.441 & 0.942  \\ 
    SimS &{5\,(c)}& 0.963 & 0.963 & 0.212 & 0.814 \\ 
    \midrule
    All  & {5\,(d)}& 0.975 & 0.975 & 0.197 & 0.799 \\
    \bottomrule
\end{tabular}	
\end{table}

\subsection{Channel Charting from Streaming CSI Data}
\label{sec:perf_comparison}

We now show channel charts obtained from the proposed methods. 
We train one Siamese neural network from $\setMr$ and one from $\setMs$ after $N=17\,516$ arrived CSI samples. 
As a baseline, we also train a Siamese network with \emph{all} $N$ arrived  samples, which we refer as ``All.''
We then evaluate the channel chart quality performance using the second trajectory from~\cite{dataset-dichasus-cf0x}; this test set consists of $23\,478$ CSI samples.

In~\fref{fig:test_charts}, we show the ground-truth positions of the UE in the test set along with three channel charts.
\fref{fig:test_charts}(b) confirms the expectation that RandoS does not perform well due to catastrophic forgetting. In fact, the training set $\setMr$ includes almost no samples from the green part of the original trajectory.
In stark contrast, Figs.~\ref{fig:test_charts}(c) and (d) obtained from SimS and the idealistic ``All'' baseline that had the full CSI dataset for training, respectively, show high-quality channel charts in which the green and red areas are well separated.

To quantify the improvement of Sims over RandoS, and to compare Sims to the baseline ``All,'' \fref{tbl:result_table} lists the associated performance metrics outlined in~\fref{sec:perf_metrics}.
As expected, RandoS performs the worst in all considered metrics and the ``All'' baseline performs best. 
Quite surprisingly, the performance of SimS with a very small core memory of only $M=1\,000$ CSI samples is at most $0.015$ worse than the ``All'' baseline in all four metrics. 
Put simply: learning a CC function from a tiny core CSI memory curated with the SimS strategy achieves comparable quality as training from \emph{all} $17\,516$ CSI samples!


\section{Conclusions and Future Work}
\label{sec:conclusions}

We have studied a realistic scenario in which CSI is estimated in a streaming fashion and the basestation has a small and fixed-capacity core memory that stores CSI data from which one learns the channel charting function. 
We have proposed two simple curation strategies to maintain the core memory: The first one stores a random subset of the streaming CSI and is mostly used to demonstrate catastrophic forgetting. The second utilizes a criterion that attempts to minimize the maximum cosine similarity between CSI samples in the core memory.
We have demonstrated with measured CSI data that our first method indeed suffers from catastrophic forgetting, while the second method results in a core memory that contains a representative subset of the streamed CSI data. 
The latter approach with a core dataset of only $1\,000$ CSI samples results in comparable channel chart quality as an idealistic approach that has access to all $17\,516$ CSI samples.

There are many avenues for future work. Other similarity metrics and more sophisticated curation strategies should be explored. 
Furthermore, the core CSI memory could store data that is more suitable for channel charting (e.g., linear combinations of arrived CSI features). 
Finally, the consistency of channel charts over time should be studied.

\bibliographystyle{IEEEtran}

\balance

\bibliography{bib/IEEEabrv,bib/confs-jrnls,bib/publishers,bib/studer,bib/vipbib, bib/cc_bib, bib/emres_bib, bib/charting} 

\begin{thebibliography}{10}
\providecommand{\url}[1]{#1}
\csname url@samestyle\endcsname
\providecommand{\newblock}{\relax}
\providecommand{\bibinfo}[2]{#2}
\providecommand{\BIBentrySTDinterwordspacing}{\spaceskip=0pt\relax}
\providecommand{\BIBentryALTinterwordstretchfactor}{4}
\providecommand{\BIBentryALTinterwordspacing}{\spaceskip=\fontdimen2\font plus
\BIBentryALTinterwordstretchfactor\fontdimen3\font minus
  \fontdimen4\font\relax}
\providecommand{\BIBforeignlanguage}[2]{{%
\expandafter\ifx\csname l@#1\endcsname\relax
\typeout{** WARNING: IEEEtran.bst: No hyphenation pattern has been}%
\typeout{** loaded for the language `#1'. Using the pattern for}%
\typeout{** the default language instead.}%
\else
\language=\csname l@#1\endcsname
\fi
#2}}
\providecommand{\BIBdecl}{\relax}
\BIBdecl

\bibitem{Studer2018}
C.~Studer, S.~Medjkouh, E.~Gönültaş, T.~Goldstein, and O.~Tirkkonen,
  ``Channel charting: Locating users within the radio environment using channel
  state information,'' \emph{IEEE Access}, vol.~6, pp. 47\,682--47\,698, Aug.
  2018.

\bibitem{ferrand2021}
P.~Ferrand, A.~Decurninge, L.~G. Ordoñez, and M.~Guillaud, ``Triplet-based
  wireless channel charting: Architecture and experiments,'' \emph{{IEEE} J.
  Sel. Areas Commun.}, vol.~39, no.~8, pp. 2361--2373, Jun. 2021.

\bibitem{ferrand2023wireless}
P.~Ferrand, M.~Guillaud, C.~Studer, and O.~Tirkkonen, ``Wireless channel
  charting: Theory, practice, and applications,'' \emph{{IEEE} Commun. Mag.},
  vol.~61, no.~6, pp. 124--130, Jun. 2023.

\bibitem{kirkpatrick2017}
J.~Kirkpatrick, R.~Pascanu, N.~Rabinowitz, J.~Veness, G.~Desjardins, A.~A.
  Rusu, K.~Milan, J.~Quan, T.~Ramalho, A.~Grabska-Barwinska \emph{et~al.},
  ``Overcoming catastrophic forgetting in neural networks,'' \emph{Proc.
  National Academy of Sciences}, vol. 114, no.~13, pp. 3521--3526, Mar. 2017.

\bibitem{dichasus2021}
F.~Euchner, M.~Gauger, S.~D\"orner, and S.~ten Brink, ``A distributed massive
  {MIMO} channel sounder for "big {CSI} data"-driven machine learning,'' in
  \emph{Proc. Int. ITG Workshop on Smart Antennas (WSA)}, Nov. 2021.

\bibitem{pmlr-v119-mirzasoleiman20a}
B.~Mirzasoleiman, J.~Bilmes, and J.~Leskovec, ``Coresets for data-efficient
  training of machine learning models,'' in \emph{Proc. Int. Conf. on Mach.
  Learn.}, vol. 119, Jul. 2020, pp. 6950--6960.

\bibitem{zhao2021condensation_gradient_matching}
B.~Zhao, K.~R. Mopuri, and H.~Bilen, ``Dataset condensation with gradient
  matching,'' \emph{arXiv preprint arXiv:2006.05929}, Mar. 2021.

\bibitem{Wievel_condensed_composite_memory_continual_learning}
F.~Wiewel and B.~Yang, ``Condensed composite memory continual learning,'' in
  \emph{Proc. Int. Joint Conf. Neural Netw. (IJCNN)}, Jul. 2021.

\bibitem{Sangermano_IJCNN22_sample_condensation_OCL}
M.~Sangermano, A.~Carta, A.~Cossu, and D.~Bacciu, ``Sample condensation in
  online continual learning,'' in \emph{Proc. Int. Joint Conf. Neural Netw.
  (IJCNN)}, Jul. 2022.

\bibitem{yu2023dataset_distillation}
R.~Yu, S.~Liu, and X.~Wang, ``Dataset distillation: A comprehensive review,''
  \emph{{IEEE} Trans. Pattern Anal. Mach. Intell.}, Oct. 2023, early access.

\bibitem{gonultas22twc}
E.~G\"{o}n\"{u}lta\c{s}, E.~Lei, J.~Langerman, H.~Huang, and C.~Studer,
  ``{CSI}-based multi-antenna and multi-point indoor positioning using
  probability fusion,'' \emph{{IEEE} Trans. Wireless Commun.}, vol.~21, no.~4,
  pp. 2162--2176, Apr. 2022.

\bibitem{dataset-dichasus-cf0x}
\BIBentryALTinterwordspacing
F.~Euchner and M.~Gauger, ``{{CSI} Dataset {dichasus-cf0x}: Distributed Antenna
  Setup in Industrial Environment, Day 1},'' 2022. [Online]. Available:
  \url{https://doi.org/doi:10.18419/darus-2854}
\BIBentrySTDinterwordspacing

\bibitem{lundpaper}
J.~{Vieira}, E.~{Leitinger}, M.~{Sarajlic}, X.~{Li}, and F.~{Tufvesson}, ``Deep
  convolutional neural networks for massive {MIMO} fingerprint-based
  positioning,'' in \emph{Proc. IEEE Intl. Symp. Personal, Indoor, Mobile Radio
  Commun.}, Oct. 2017, pp. 1--6.

\bibitem{lei19siamese}
E.~Lei, O.~Castañeda, O.~Tirkkonen, T.~Goldstein, and C.~Studer, ``Siamese
  neural networks for wireless positioning and channel charting,'' in
  \emph{Proc. Allerton Conf. Commun., Contr., Comput.}, Sep. 2019, pp.
  200--207.

\bibitem{taner2023globecom}
S.~Taner, V.~Palhares, and C.~Studer, ``Channel charting in real-world
  coordinates,'' presented at the IEEE Global Telecommun. Conf. (GLOBECOM),
  Dec. 2023.

\bibitem{penghzipaper}
P.~{Huang}, O.~{Casta{\~n}eda}, E.~{G{\"o}n{\"u}lta{\c s}}, S.~{Medjkouh},
  O.~{Tirkkonen}, T.~{Goldstein}, and C.~{Studer}, ``Improving channel charting
  with representation-constrained autoencoders,'' in \emph{Proc. IEEE Int.
  Workshop Signal Process. Advances Wireless Commun. (SPAWC)}, Aug. 2019, pp.
  1--5.

\bibitem{yassine22}
T.~Yassine, L.~L. Magoarou, S.~Paquelet, and M.~Crussière, ``Leveraging
  triplet loss and nonlinear dimensionality reduction for on-the-fly channel
  charting,'' in \emph{Proc. IEEE Int. Workshop Signal Process. Advances
  Wireless Commun. (SPAWC)}, Jul. 2022.

\bibitem{rappaport2021}
B.~Rappaport, E.~Gönültaş, J.~Hoydis, M.~Arnold, P.~K. Srinath, and
  C.~Studer, ``Improving channel charting using a split triplet loss and an
  inertial regularizer,'' in \emph{Int. Symp. on Wireless Commun. Sys.
  (ISWCS)}, Sep. 2021.

\bibitem{stahlke23}
M.~Stahlke, G.~Yammine, T.~Feigl, B.~M. Eskofier, and C.~Mutschler, ``Indoor
  localization with robust global channel charting: A time-distance-based
  approach,'' \emph{{IEEE} Trans. on Mach. Learn. in Commun. and Netw.}, Mar.
  2023, early access.

\bibitem{stephan2023adp}
P.~Stephan, F.~Euchner, and S.~ten Brink, ``Angle-delay profile-based and
  timestamp-aided dissimilarity metrics for channel charting,'' \emph{arXiv
  preprint arXiv:2308.09539}, Sep. 2023.

\bibitem{dijkstra1959note}
E.~W. Dijkstra, ``A note on two problems in connexion with graphs,''
  \emph{Numerische mathematik}, vol.~1, no.~1, pp. 269--271, 1959.

\bibitem{glorot}
X.~Glorot and Y.~Bengio, ``Understanding the difficulty of training deep
  feedforward neural networks,'' in \emph{Proc. Thirteenth Int. Conf.
  Artificial Intelligence Stat.}, vol.~9, May 2010, pp. 249--256.

\bibitem{kingma2014adam}
D.~P. Kingma and J.~Ba, ``Adam: A method for stochastic optimization,''
  \emph{arXiv preprint arXiv:1412.6980}, Jan. 2017.

\bibitem{altous22asilomar}
H.~Al–Tous, P.~Kazemi, C.~Studer, and O.~Tirkkonen, ``Channel charting with
  angle-delay-power-profile features and earth-mover distance,'' in \emph{Proc.
  Asilomar Conf. Signals, Syst., Comput.}, Oct. 2022, pp. 1195--1201.

\end{thebibliography}

\balance
	
\end{document}